# DISCRETE ELEMENTS METHOD: A NEW KIND OF INITIAL CONDITIONS –TETRAHEDRAL PACKING OF BALLS[1]

## Mark A.Tsayger[2]

The Paper discusses the use of the regular packing of identical balls with the coordination number 4 as a model of a medium consisting of fluid and solid particles in the conditions of fluidization. It is proposed to use the examined packing of balls as an initial condition for the calculations by the Discrete Elements Method (DEM) in technological processes in the fluidized bed in industry, as well as in the modeling of processes that occur in hydraulic and pneumatic transport of granular materials. Filtration properties of such packing required for its use as an initial condition in calculations by DEM are estimated.

In the past several decades, the method of discrete elements (Discrete Element Method - DEM) has become widespread in industrial chemical engineering calculations. This method of modeling processes involving granular media originally proposed by Cundall [1] considers the behavior of particles under the influence of various forces, mainly the forces of contact interaction, and the impact forces of the surrounding fluid. In DEM method, particles are replaced with elastic balls of the same diameter. The development of computer technology has made it possible to perform iterative computations for each of the plurality of particles numbering up to tens thousands and even more.

In fact, DEM method is a kind of a method of successive approximations whose results are essentially independent of the initial conditions. However, different methods of successive approximations have different convergences, and the choice of initial conditions affects the labor intensiveness of computations – initial conditions which are closer to the reality consume less labor.

Both regular and irregular packings of balls can be taken as initial conditions. The advantage of regular packing consists in known initial coordinates of positions of balls, which facilitated and simplified the calculation of force interactions, including those of ball interaction with the surrounding fluid.

There are three types of regular packings of balls with different coordination numbers – the number of contacts of a ball with surrounding balls. Packing with coordination number 12 is called hexagonal, packing with a coordination number 6 is called cubic. Both of these packings were examined by Slichter [2]. Another regular packing with the coordination number 4 is known. In this packing every ball is in contact with four surrounding balls, and every ball is the center of a regular tetrahedron whose four vertices are centers of the contacting balls. Hilbert, who has studied this packing and calculated its porosity [3, pp. 48-50] calls it tetrahedral packing. Such packing is observed in the crystal lattice of silicon, diamond and myriads of other substances (Fig. 1)[3]. From the author's knowledge, this package has not been considered earlier as an initial condition for the implementation of DEM.

[2] Mark A.Tsayger, PhD, Beer Sheva, Israel. E-mail: m_tsayger@hotmail.com

To use the packing of balls as the initial condition, it is desirable to know its characteristics, such as porosity and hydraulic resistance to the fluid flow through the packing. Porosity is a fraction of space surrounding the balls in a unit cell of packing with respect to the entire space of the unit cell (which includes interstitial spaces and balls themselves). In contrast to the well-known

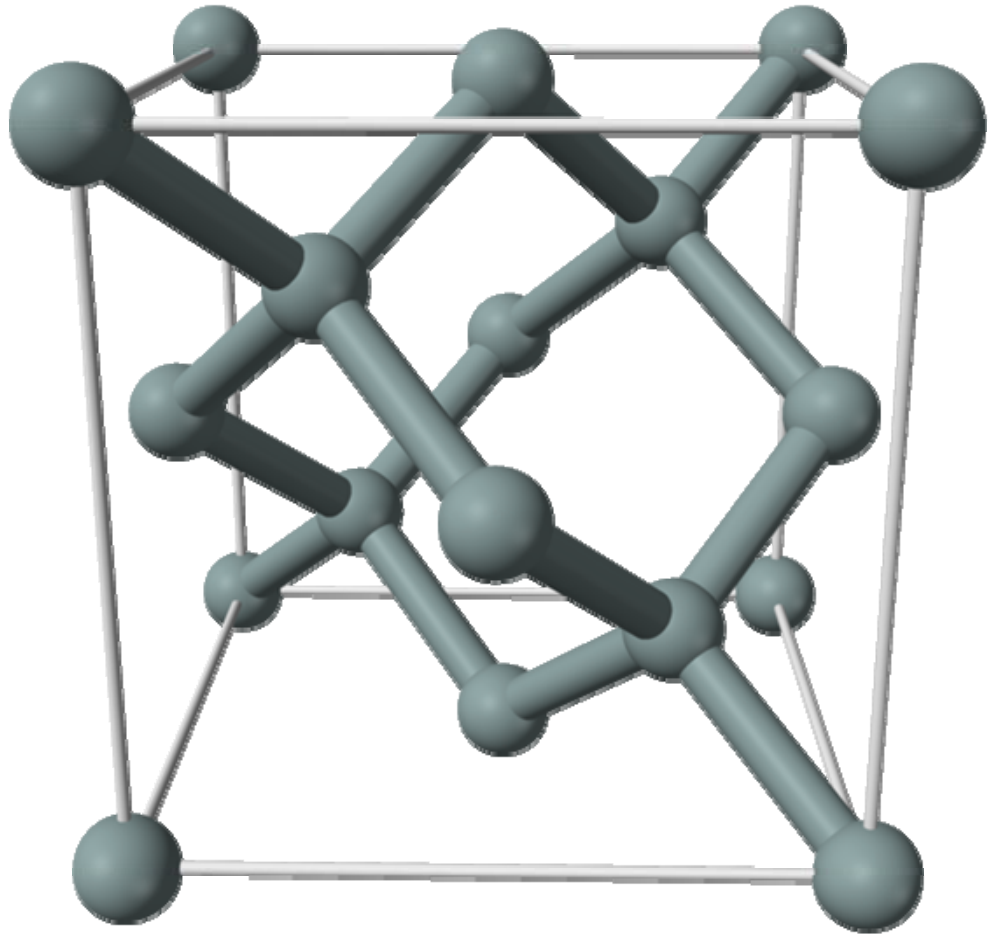

Fig 1. Silicon lattice (cubic, face-centered). Single-crystal Si
(http://en.wikipedia.org/wiki/File:Silicon-unit-cell-3D-balls.png)

---

[3] Author intentionally uses this well-known figure despite the fact that it shows balls of reduced diameter, which do not touch each other. But this figure clearly shows relative positions of balls in the unit cell, which is necessary to calculate the porosity of the structure.

Hilbert's calculation of the porosity of regular packing with the coordination number 4, a more visual calculation yielding the known result is given below. This technique has allowed us to calculate the coordinates of balls in a unit cell of the packing, which are given below in Table 1.

From the data about a regular tetrahedron, the radius $R$ of a sphere circumscribed around a regular tetrahedron with an edge $a$ is

$$R = \frac{a\sqrt{6}}{4}$$

In our case, the radius of the circumscribed sphere is exactly equal to the diameter $d$ of balls of the packing, so the edge of a regular tetrahedron is equal to

$$a = \frac{2}{3}\sqrt{6}\, d$$

As Fig. 1 shows, the diagonal of a square face of the packing unit cell is $2\,a$, wherefrom we obtain the length of the edge of the packing unit cell $l$ equal to

$$l = \frac{4}{\sqrt{3}}\, d$$

Consequently, the volume of the unit cell of the regular tetrahedral packing is

$$l^3 = \frac{64}{3\sqrt{3}}\, d^3$$

Now we count how many balls of the regular tetrahedral packing are located in a single unit cell. Figure 1 shows that in each of eight angles of the unit cell there is one ball, so that 1/8 of the ball fits inside the unit cell at each angle, which amounts to a total of 1 ball. Further, there is one ball in the center of each of the six square faces of the unit cell, that is, half of the ball fits inside the cell in each of face-centered points, making a total of 3 balls. Finally, 4 balls that are centers of four regular tetrahedrons are completely within the cell. Thus, 8 balls are located inside the unit cell of the tetrahedral packing of fictitious soil. Their volume $V_s$ is

$$V_s = \frac{4}{3}\pi d^3$$

From the definition of porosity $\varepsilon$ as a ratio of the unit cell space surrounding the balls to the total volume of the cell, we obtain

$$\varepsilon = \frac{l^3 - V_s}{l^3} = 1 - \frac{\sqrt{3}}{16}\pi = 0.65991\ldots$$

i.e., for the regular tetrahedral packing, the porosity is ~0.660. The porosity for Slichter's hexagonal packing is ~0.259, while for Slichter's cubic packing we have $\varepsilon = \sim 0.476$. In other words, the regular tetrahedral packing of balls has a higher porosity than that of the loosest cubic packing offered by Slichter.

While the porosity of the tetrahedral packing is relatively easy to calculate, it is not so easy to calculate the filtration characteristics of such a packing. Despite numerous experimental studies of various kinds of granular fillings, including beds represented by balls of the same diameter, the author is unaware of any researches that ensure a proper compliance to the regular nature of packing throughout the whole volume of a porous bed during the experiment. There are many studies of filtration characteristics of fillings with balls of the same diameter and a random structure of packing. If we consider Slichter's cubic packing (coordination number = 6), we can see that without special measures to ensure such nature of packing of balls, the porous bed can be repacked into a denser structure during the experiment, which will affect the results of measurements of filtration characteristics of the bed. And for the regular tetrahedral packing schematically shown in Fig. 1, its instability is apparent even by eye, and without special measures for conserving the character of packing in the sample, experimental measurements of filtration characteristics of this packing are impossible. Special measures to comply with the regular character of balls packing in the case of cubic packing are presented as thin rigid pins connecting every two neighboring balls in their contact point, i.e., each ball must be connected to the adjacent balls with six pins (which is the coordination number for a cubic packing). Measures ensuring the rigidity of a tetrahedral packing of balls represent a complicated engineering problem, whose solution is not obvious.

However, the evaluation of filtration resistance of a regular tetrahedral packing is vital when this packing is used as the initial condition in the calculations of granular medium behavior in the studied processes, where it exists jointly with a moving fluid, by the DEM.

The author has decided to obtain the filtration resistance of the regular tetrahedral packing *in the first approximation* using the results of experimental studies of hydraulic resistance of balls at their constrained falling in fluid.

It is known [4-6] that the velocity $w$ of a spherical body fall in a fluid depends on the diameter of the ball $d$, gravitational acceleration $g$, density of the ball $\Delta$, fluid density $\rho$ and viscosity $\mu$, the criterial dependence $Re = f(Ar)$ being represented by a unified experimental curve in the whole range of parameters at laminar, turbulent and intermediate flow patterns with an accuracy sufficient for practical purposes up to the values $Ar \approx 10^7$ (see [4-6]). Here $Re = \dfrac{v\,d\,\rho}{\mu}$ is the Reynolds criterion,

$$Ar = \frac{g d^3}{\mu^2}\rho(\Delta - \rho) - \text{the Archimedes criterion}$$

At the settling of a large number of identical grains or at their suspension in an ascending flow of liquid or gas, the velocity of their constrained falling or the suspending flow velocity $w$ are determined not only by the Archimedes criterion, but also by the volume fractions of both phases. Denote a fraction of the free volume between the balls (or the porosity of the layer of balls) by $\varepsilon$. Then the volume fraction of the balls themselves is $(1 - \varepsilon)$.

In a number of experimental studies, whose bibliography is presented in [4-6], close interpolation binomial formulas with slightly different numerical coefficients were obtained. As shown by Ergun [4-5], data of various authors are well described by the equation in the form transformed by Goroshko et al. [6]:

$$Ar = 150 \frac{1-\varepsilon}{\varepsilon^3} Re_c + 1.75 \frac{1}{\varepsilon^3} Re_c{}^2 \qquad (1)$$

where $Re_c$ is the Reynolds criterion under critical conditions, when the layer of balls starts passing from a motionless into a suspended state. The Reynolds criterion is obtained using an superficial fluid velocity $w$ calculated for the total cross section of the flow, which includes the space around the balls and the space occupied by the balls.

**The author assumes that in these critical conditions (i.e., when the layer of balls is ready to lose immobility), the relative positions of the balls in a suspended layer approximately correspond to a regular tetrahedral packing.**

Therefore, equation (1) is used for an approximate estimate of filtration characteristics of the tetrahedral type of packing. Substituting as porosity value $\varepsilon \approx 0.660$, derived above, we obtain for the tetrahedral packing:

$$Ar = 177.394\, Re + 5.087\, Re^2 \qquad (2)$$

The author assumes that equation (1) is also consistent with Slichter's cubic and hexagonal packings. Then for the cubic packing of Slichter ($\varepsilon \approx 0.476$) we have:

$$Ar = 728.79\, Re + 16.226\, Re^2 \qquad (3)$$

For the densest Slichter's hexagonal packing ($\varepsilon \approx 0{,}259$) we have

$$Ar = 6397.5\, Re + 100.7\, Re^2 \qquad (4)$$

These assumptions enable us to get a rough idea of how the permeability of a regular packing of balls increases at the transition from hexagonal to cubic and, further, to tetrahedral packing.

The permeability can be roughly considered as reciprocal to the coefficient at the first power of the Reynolds criterion in equations (2 ÷ 4). Taking the permeability of hexagonal packing as 1, the transition to cubic packing results in a 9.8-fold increase in permeability, and for the loosest tetrahedral packing the permeability grows up to a 36-fold value.

Of course, so far these are only assumptions, but they are rather intriguing and, quite possibly, not far for the truth, if we take for the truth the results of special studies of hydraulic resistance of various types of regular packings of balls, with special measures for the preservation of the packing pattern during the experiments. In our opinion, it is useful to carry out such special studies in the future.

Taking into account that in DEM method a slight inaccuracy in the initial conditions has a minimal impact on the duration and complexity of calculations, the author believes that using the tetrahedral packing as the initial condition for the calculation by DEM method, equation (2) can be used to find the initial velocity of the fluid with respect to fixed balls.

Using a regular tetrahedral packing of balls as the initial condition requires a proper initial placement of balls in the modeled area. The coordinates of the centers of balls can be calculated using an auxiliary Table 1.

Table 1 shows the location of eight balls of the unit cell relative to the zero point. The diameter of the ball is accepted as a unit of length. Since all the space of regular packing of balls can be obtained by doubling the placement of balls in the unit cell, Table 1 allows us to set the initial coordinates of the balls at any point of the modeled space.

| Nominal index of the ball | Coordinate x | Coordinate y | Coordinate z |
|---|---|---|---|
| A | 0 | 0 | 0 |
| B | $\dfrac{2}{\sqrt{3}}d$ | $\dfrac{2}{\sqrt{3}}d$ | 0 |
| C | $\dfrac{2}{\sqrt{3}}d$ | 0 | $\dfrac{2}{\sqrt{3}}d$ |
| D | 0 | $\dfrac{2}{\sqrt{3}}d$ | $\dfrac{2}{\sqrt{3}}d$ |
| E | $\dfrac{1}{\sqrt{3}}d$ | $\dfrac{3}{\sqrt{3}}d$ | $\dfrac{3}{\sqrt{3}}d$ |
| F | $\dfrac{3}{\sqrt{3}}d$ | $\dfrac{1}{\sqrt{3}}d$ | $\dfrac{3}{\sqrt{3}}d$ |
| G | $\dfrac{1}{\sqrt{3}}d$ | $\dfrac{1}{\sqrt{3}}d$ | $\dfrac{1}{\sqrt{3}}d$ |
| H | $\dfrac{3}{\sqrt{3}}d$ | $\dfrac{3}{\sqrt{3}}d$ | $\dfrac{1}{\sqrt{3}}d$ |

Notes: 1) the ball diameter is $d$ ; 2) the unit cell is a cube with the edge $\dfrac{4}{\sqrt{3}}d$ ; 3) the distance between points E and D is d; 4) the distance between points F and C is d; 5) the distances between points G and A, as well as between points G and B and between points G and C and between points B and H are equal to d.

Table 1. The ball coordinates of regular tetrahedral packing in the unit cell

Example: Calculate the coordinates of the center of a ball in a regular tetrahedral packing. It is a ball with a nominal index E in the unit cell which is located at the fifth position in X-direction, in the sixth row in Y-direction and in the third layer in Z-direction. The initial unit cell of packing is located at the zero position, in the zero row and zero layer.

Solution: Determine the coordinates of the zero point A of the considered unit cell.

$$X_A = 5 * \frac{4}{\sqrt{3}}d = \frac{20}{\sqrt{3}}d$$

$$Y_A = 6 * \frac{4}{\sqrt{3}}d \ = \frac{24}{\sqrt{3}}d$$

$$Z_A = 3 * \frac{4}{\sqrt{3}}d = \frac{12}{\sqrt{3}}d$$

Determine the coordinates of the required center of the ball with the index E using Table 1.

$$X_E = \frac{20}{\sqrt{3}}d \ + \frac{1}{\sqrt{3}}d = \frac{21}{\sqrt{3}}d$$

$$Y_E = \frac{24}{\sqrt{3}}d + \frac{3}{\sqrt{3}}d = \frac{27}{\sqrt{3}}d$$

$$Z_E = \frac{12}{\sqrt{3}}d + \frac{3}{\sqrt{3}}d = \frac{15}{\sqrt{3}}d$$

In Appendix given the presentation "A tetrahedral packing of balls.pptx" (authors M.Tsayger and L.Baluashvili), which shows a block of balls with regular tetrahedral packing, rotating at different angles. The block comprises two unit cells along the length, width and height. One can see the presence of holes that permeate the entire structure. These holes defined by features of the regular tetrahedral packing.

The initial values of the apparent average fluid velocity $w$ with respect to an immovable ball are determined by the solution of a quadratic equation (2), since at the initial moment of time all the parameters of the equation except the velocity $w$ are known.

Initial conditions in the form of a tetrahedral packing are proposed for calculating the processes in which a granular material at the initial point of the process is in the state of fluidization (this may be a variety of two-phase processes in a fluidized bed, processes of two-phase pneumatic or hydraulic transport of granular materials, etc.)

The proposed scheme for calculating initial conditions using a regular tetrahedral packing of balls is useful for calculations of the behavior of granular fillings under the conditions of fluidization, when flow resistance of the filling is equal to forces acting on filling upwards, using the discrete element method (DEM). Such calculations can be aimed at the study of the behavior of granular media in technological processes in a fluidized bed, as well as at the transfer of granular materials by pipelines. Application of the proposed scheme can greatly simplify and reduce the cost of calculations performed by DEM.

## ACKNOWLEDGMENTS

The author would like to thank Mrs. Natalia Goldbaum for her assistance in preparing this Paper.

APPENDIX

M.Tsayger and L.Baluashvili *A tetrahedral packing of balls*.pptx

Cloud storage:

https://drive.google.com/file/d/0BzHdUn9T2yJRN2g1ZmJaWlN6M0U/view?usp=sharing

Presentation size 275 MB